\documentclass[twocolumn, amsmath, nohypertex, amsfonts,prb]{revtex4}
\usepackage{graphicx}
\usepackage{url}
\pdfoutput=1
\begin{document}

\title{Adiabatic theory of SET and RESET transitions}

\author{V. G. Karpov}
\email{victor.karpov@utoledo.edu}
\affiliation{Department of Physics and Astronomy, University of Toledo, Toledo, OH 43606, USA}

\date{\today}

\begin{abstract}

We develop a phenomenological theory of pulse induced phase transformations behind the SET (from high to low resistive state) and RESET (backward) processes in nonvolatile memory. We show that in modern era devices, both evolve in the adiabatic regime with energy deposition time much shorter than that of thermalization. They are however different by the operating modes:  voltage source driven for SET and current source driven for RESET. The characteristic temperatures and transition rates are expressed through material and process parameters.
\end{abstract}
\maketitle

\section{Introduction} \label{sec:intro}
SET and RESET processes underly operations of phase change memory (PCM) and resistive random access memory (RRAM). They switch a system respectively from its high resistive state to the low one and vise versa. The switchings occur through local structural transformations between insulating and conductive phases, microscopically different between different materials. \cite{wong2012,raoux2014,PCM,RRAM} Here we develop a phenomenological approach that is not limited to any particular microscopic structures describing SET and RESET in terms of power generation and heat transfer.

\begin{figure}[b!]
\includegraphics[width=0.49\textwidth]{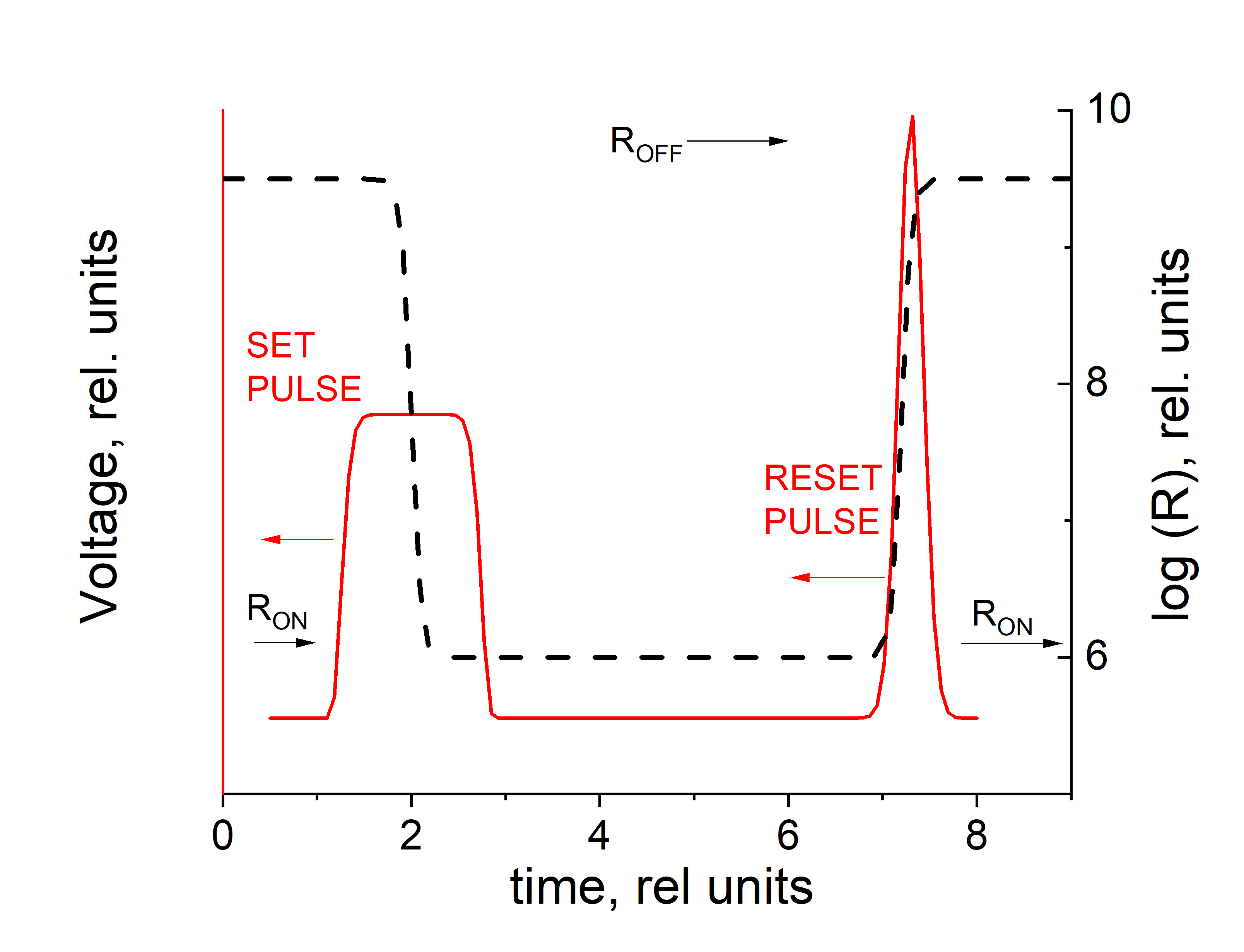}
\caption{A generic sketch of temporal dependencies of device voltage (solid line) and resistance (dash) showing SET and RESET pulses and OFF, ON resistances. \label{Fig:ONOFF}}
\end{figure}

We recall that switching from the high to low resistance state is triggered by SET pulses, while the RESET ones initiate the reversal process. As illustrated in Fig. \ref{Fig:ONOFF}, it is typical that SET pulses have lower amplitudes and longer durations than the RESET ones. The general thermodynamic description of both processes is given by the heat transfer equation treated in what follows in the adiabatic approximation.

An important difference between SET and RESET processes is that the former starts from the high resistance (OFF) state with the switching region resistance $R_{\rm OFF}$ exceeding that of other circuitry elements in series. The latter inequality implies a more or less fixed voltage across $R_{\rm OFF}$; hence, voltage source driven regime. To the contrary, RESET starts with the low resistance of the switching region (ON state). For $R_{\rm ON}$ lower than the resistance of other circuitry elements in series, the switching region finds itself under approximately fixed current $I$; hence, current source driven regime.

The latter distinction between the voltage and current driven RESET and SET processes was not systematically explored, although it was commonly recognized however that SET and RESET transitions are better characterized by respectively threshold voltages and threshold currents. Also, it was noticed \cite{karpov2016,karpov2017} that the current voltage dependencies with the characteristic domains of almost constant $U$ for the SET and almost constant $I$ for RESET regimes, could be derived from the free energy minimum under the conditions of $U=const$ and $I=const$ respectively.

Here we show how the voltage vs. current driven processes determine the differences between SET and RESET kinetics allowing simple analytical descriptions. Our paper is organized as follows. Sec. \ref{sec:ht} introduces the adiabatic approximation for SET and RESET heat transfer equations describing their temporal temperature dependencies. Sec. \ref{sec:st} shows how those dependencies define the SET and RESET transition kinetics. The opposite to the adiabatic, quasi-stationary regime is analyzed in Sec.  \ref{sec:qs} The discussion in Sec. \ref{sec:disc} relates the obtained results to some observations. The conclusions are given in Sec. \ref{sec:concl}.

\section{Heat transfer and adiabatic regime}\label{sec:ht}
We assume, on empirical grounds, \cite{wong2012,raoux2014,PCM,RRAM,kim2014,savransky2008,mechonic2012,tang2008} temperature activated resistances,
\begin{equation}\label{eq:ROFF}R_{\rm OFF(ON)}=R_0\exp\left(\frac{W_{\rm OFF(ON)}}{kT}\right).
\end{equation}
Then the power generated in the OFF and ON states becomes respectively $P=U^2/R_{\rm OFF}$ and  $P=I^2R_{\rm ON}$ with opposite temperature dependencies,
\begin{eqnarray}\label{eq:power} P&=&P_0\exp\left(-\frac{W_{\rm OFF}}{kT}\right), \quad P_0\equiv \frac{U^2}{R_0}\quad {\rm - SET}, \label{eq:PR}\\
P&=&P_0\exp\left(\frac{W_{\rm ON}}{kT}\right),\quad P_0\equiv I^2R_0 \quad {\rm - RESET}\label{eq:PS}.\end{eqnarray}
$P_0$ can be a function of time governed ny the voltage or current pulse shapes $U(t)$, $I(t)$.
Experimentally, for various materials, $W_{\rm OFF}$ is on the order of 1-0.5 eV, while $W_{\rm ON}$ is two or more times lower than $W_{\rm OFF}$, remaining much greater than the thermal energy $kT$. \cite{wong2012,raoux2014,kim2014,savransky2008,mechonic2012,tang2008}   We will separately consider the case \cite{govorenau2013} of $W_{\rm ON}=0$.

Note that the origin of thermally activated resistances in Eq. (\ref{eq:ROFF}) is not specified here (various possibilities are discussed in Refs. \cite{wong2012,raoux2014,PCM,RRAM,rev}). Our description here is purely phenomenological and applies to all possible mechanisms of activated conduction.

Our treatment of both the SET and RESET processes is based on the general heat transfer equation,
\begin{equation}\label{eq:HT}
c\rho\frac{\partial T}{\partial t}=\frac{P}{a^3}+\chi \nabla ^2 T,
\end{equation}
where $c$ and $\rho$ are respectively the specific heat and material density, $a^3$ is the characteristic volume, in which that power is dissipated, and $\chi$ is the thermal conductivity. The characteristic spatial dimension $a$ is in the nanometer region for the modern technology devices. Eq. (\ref{eq:HT}) simplifies in the adiabatic regime where the heat diffusivity can be neglected. To justify its applicability, we define two characteristic times as described next.

The heating time $\tau _H$ is estimated by approximating $\partial T/\partial t\sim \delta T/\tau _H$, i. e. $\tau _H\sim c\rho a^3\delta T/P$ where $\delta T$ is the temperature increase.
On the other hand, the thermal diffusivity time $\tau _D$ can estimated in the approximation $\partial T/\partial t\sim \delta T/\tau _D$ and $\nabla ^2T\sim \delta T/a^2$ implying thermal diffusion over distances of the order of $a$. This yields the well known result, $\tau _D\sim a^2c\rho /\chi $.

We define the adiabatic parameter
\begin{equation}\label{eq:alpha}
\alpha \equiv \tau _H/\tau _D\sim \chi a\delta T/ P.\end{equation}
When $\alpha\ll 1$, heat is confined to the local region of length $a$ and can be described adiabatically, neglecting thermal exchange.

Before considering the adiabatic description, we give some estimates for $\alpha$. Assuming a single length scale $a$ for the active region, its resistance $R\sim r/a$ where $r$ is the resistivity. Representing the power $P\sim U^2a/r$ where $U$ is the voltage across active region, yields $\alpha \sim (r\chi) \delta T/U^2$. The latter form is more convenient because it does not include linear dimension $a$. For metals, the Wiedemann-Franz law predicts $(r\chi )\sim (k/e)^2T $, i. e. $\alpha \sim k^2T\delta T/(eU)^2$, which can be applicable to the ON state. Because $eU\sim 1$ eV and $kT\lesssim 0.1 eV$, we obtain $\alpha \ll 1$.

For the insulating OFF state, the Wiedemann-Franz relation can still apply as a `rule of thumb' if both  thermal and electric transports are dominated by the electrons. \cite{mahan1999,karpov2019} Otherwise, with $r$ and $\chi$ for various systems, the product $r\chi (e/k)^2/T$ is still not large enough to violate the adiabatic criterion $\alpha\ll 1$.  However, for completeness, we will describe as well the alternative case of $\alpha \sim 1$ in Sec. \ref{sec:qs}.

In the adiabatic approximation, $\alpha\ll 1$, the second term on the right-hand-side of Eq. (\ref{eq:HT}) is neglected  and temperature $T(t)$ after time $t$ since heating from the initial temperature $T_i$ is found from the integral,
\begin{equation}\label{eq:int}
\int _{T_i}^T\frac{P_0dT}{P(T)}=\frac{1}{c\rho a^3}\int _0^t P_0(t)dt\equiv \frac{A(t)}{c\rho a^3}. \end{equation}
For the RESET and SET processes, the left-hand-side integrand represents respectively $\exp(-W_{\rm OFF}/kT)$ and $\exp(W_{\rm ON}/kT)$, while the right-hand-side $A$ (correlated with injected energy) can be evaluated using $P_0$ from Eqs. (\ref{eq:PR}), (\ref{eq:PS}) along with applied pulse shapes and empirically estimated $R_0$.

Because of the opposite sign exponents in their integrands, the left-hand-side integrals in Eq. (\ref{eq:int}) are estimated differently for the RESET and SET processes. The former is determined by the highest temperature region around its upper limit, while the latter is dominated by the proximity of its lower limit.

We approximate the RESET case integral with the product of its integrand width and amplitude, reducing Eq. (\ref{eq:int}) to the form,
$(kT^2/W_{\rm ON})\exp(-W_{\rm ON}/kT)=A/\rho ca^3 $. In addition, we take into account that $c\sim k/m$ where m is the characteristic atomic mass, and $\rho\sim m/a_0^3$ where $a_0$ is the characteristic interatomic distance in the material.  Therefore, to the accuracy of a numerical multiplier, one can estimate $\rho a^3 c\sim Nk$ where $N (\gg 1)$ is the number of atoms in the region of diameter $a$. For realistic parameters (see below), $W_{\rm ON}N/A\gg W_{\rm ON}/kT\gg 1$. As a result, the expression for RESET temperature simplifies to the form,
\begin{equation}\label{eq:RESET} T=W_{\rm ON}/[k\ln (W_{\rm ON}N/A)].\end{equation}

Along the same lines [and following the linearization of exponent recipe, \cite{FK,LL} $T=T_{\rm SET}-\Theta$ with $\Theta\ll T_{\rm SET}$], the expression for SET temperature becomes,
\begin{equation}\label{eq:SET}T=T_i-(T_i^2/W_{\rm OFF})\ln[1-A/(W_{\rm OFF}N)].\end{equation}
Keeping in mind that $A=A(t)$, Eqs. (\ref{eq:RESET}) and (\ref{eq:SET}) describe the temporal dependencies of RESET and SET processes.

Three comments are in order here. First, Eq. (\ref{eq:SET}) is similar to that describing the adiabatic processes in chemical kinetics. \cite{FK} Secondly, its applicability is limited to the condition $A/(W_{\rm OFF}N)\ll 1$; the description beyond that condition, can be developed along the lines of Ref. \cite{FK} Thirdly, while the linearization of exponent recipe is not explicitly validated here it is known to be at least semi-quantitatively accurate in similar problems of chemical kinetics \cite{FK} and as such included with a rigorous collection of methods. \cite{LL}

\section{Transition rates}\label{sec:st}

The very fact that ON and OFF (nonvolatile memory) states are long lived means that their related SET and RESET transitions are suppressed by certain energy barriers; we denote them respectively $w_{\rm S}$ and $w_{\rm R}$. These barriers correspond to structural transformations manifesting themselves in differences between the two states. Such transformations are triggered by the external electric stimuli, which alter the barriers lowering them enough to allow the desired transition rates. As long as the barriers remain finite they will be surmounted by thermal activations; hence, transition rates proportional to $\exp[-w_{{\rm R(S)}}/kT]$ with the barriers $w_{\rm R}$ and $w_{\rm S}$ reduced by the stimuli.

The above introduced activation energies $w_{\rm R}$ and $w_{\rm S}$ do not specify any microscopic mechanisms behind the RESET and SET processes. They can be electric field dependent, as e. g. the field induced nucleation energy of conductive filaments, \cite{karpov2017} or represent a melting process enthalpy. Eqs. (\ref{eq:RESETtr}) and (\ref{eq:SETtr}) will hold for all conceivable mechanisms as long as they satisfy the condition of adiabaticity. Our description remains purely phenomenological. It applies equally to PCM and various RRAM devices: conductive bridge RAM, oxide-based RA, etc.

Here we consider the system evolution upon the stimuli application when $w_{\rm R}$ and $w_{\rm S}$ are already reduced. We consider the system evolution due to thermally activated structure rearrangements. That evolution will be strongly affected by the temporal temperatures dependencies derived in Sec. \ref{sec:ht}.

More specifically, the temporal dependencies in Eqs. (\ref{eq:RESET}) and (\ref{eq:SET}) can be accounted for in the rates of RESET and SET transformations with time dependent temperature $T(t)$,
\begin{equation}\label{eq:trans}\frac{dn_{\rm R}}{dt}=\Gamma _{\rm R}\exp\left(-\frac{w_{\rm R}}{kT}\right) \end{equation}
and similar for SET transitions. Plugging in the dependence from Eq. (\ref{eq:RESET}) yields
\begin{equation}\label{eq:RESETtr}\frac{dn_{\rm R}}{dt}=\Gamma _{\rm R}\left(\frac{A(t)}{W_{\rm ON}N}\right)^
\gamma , \ \gamma \equiv \frac{w_{\rm R}}{W_{\rm ON}}. \end{equation}
Similarly, the dependence from Eq. (\ref{eq:SET}) yields
\begin{equation}\label{eq:SETtr}\frac{dn_{\rm S}}{dt}=\Gamma _{\rm S}\exp\left(-\frac{w_{\rm S}}{kT_i}\right)\exp\left(\frac{Aw_{\rm S}}{W_{\rm OFF}^2N}\right).\end{equation}

As a relevant example, we assume $A(t)$ of the type shown in Fig. \ref{Fig:ONOFF}: $A=pt$ for SET and $A=st^2$ for the RESET process where $p$ and $s$ are two constants. Such $A(t)$ corresponding respectively to the conditions of constant voltage and linearly ramped current, yield,
\begin{eqnarray}&&\frac{dn_{\rm R}}{dt}=\Gamma _{\rm R}\left(\frac{st^2}{W_{\rm ON}N}\right)^\gamma ,\\
&&\frac{dn_{\rm S}}{dt}=\Gamma _{\rm S}\exp\left(-\frac{w_{\rm S}}{kT_i}\right)\exp\left(\frac{t}{\tau _{ind}^S}\right),
\end{eqnarray}
where we have introduced the SET induction time
\begin{equation}\label{eq:tind} \tau _{ind}^S=\frac{W_{\rm OFF}^2N}{pw_{\rm S}}.\end{equation}

Another representation of the latter results refers to the characteristic process times $\tau _{\rm R(S)}$ derived from the condition $dn_{\rm R(S)}/dt=1/\tau _{\rm R(S)}$. Straightforward calculations yield,
\begin{equation}\label{eq:tRESET}\tau _{\rm R}=\left[\frac{2\gamma +1}{\Gamma _R}\left(\frac{W_{\rm ON}N}{s}\right)^{\gamma}\right]^{1/(2\gamma +1)},\end{equation}
and
\begin{eqnarray}\label{eq:S1}\tau _{\rm S}&=&\frac{1}{\Gamma _S}\exp\left(\frac{w_{\rm S}}{kT_i}\right) \quad {\rm when}\quad  \tau _{\rm S} \ll \tau _{ind}^S,\\
\tau _{\rm S}&=&\tau _{ind}\left[ \frac{w_{\rm S}}{kT_i}-\ln (\Gamma _S\tau _{ind}^S)\right]\quad {\rm when}\quad  \tau _{\rm S} \gg \tau _{ind}^S.\nonumber
\end{eqnarray}

We separately note the case of zero (or very small) activation energy $W_{\rm ON}$, in which $P=P_0$ and Eq. (\ref{eq:int}) yields the time dependent temperature
\begin{equation}\label{eq:zeroW}T=T_i+A(t)/(Nk).\end{equation}
Substituting the latter into Eq. (\ref{eq:trans}) will describe the RESET kinetics for zero $W_{\rm ON}$. In particular, in the above mentioned approximation \cite{FK} of linearized activation exponent, one gets,
\begin{equation}\label{eq:rate0}\frac{dn_{\rm R}}{dt}=\Gamma _{\rm R}\exp\left(-\frac{w_R}{kT_i}\right)\exp\left(\frac{t}{\tau _{ind}^R}\right)^2
\end{equation}
with RESET induction time,
\begin{equation}\label{eq:tindR} \tau _{ind}^R=\sqrt{\frac{N (kT_i)^2}{sw_{\rm R}}}.\end{equation}
The corresponding process time is given by
\begin{equation}\label{eq:tR}\tau _{\rm R}=\tau _{ind}^R\sqrt{\frac{w_{\rm R}}{kT_i}-\ln (\Gamma _R\tau _{ind}^R)} \end{equation}
instead of Eq. (\ref{eq:tRESET}).

Finally, we note that other temporal dependencies $A(t)$ can be readily implemented with the above analysis. For example, assuming ramping set current $A=s't^2$ for Eq. (\ref{eq:SETtr}) yields
\begin{equation}\label{eq:Sramp}\frac{dn_{\rm S}}{dt}=\Gamma _{\rm S}\exp\left(-\frac{w_{\rm S}}{kT_i}\right)\exp\left(\frac{t}{\tau _{ind}^{S1}}\right)^2\end{equation} with the induction and process times being respectively
\begin{equation}\label{eq:tind1} \tau _{ind}^{S1}=\sqrt{\frac{W_{\rm OFF}^2N}{s'w_{\rm S}}}, \ \tau _{\rm S}^1=\tau _{ind}^R\sqrt{\frac{w_{\rm S}}{kT_i}-\ln (\Gamma _S\tau _{ind}^{S1})}.\end{equation}

\section{Quasi-stationary regime}\label{sec:qs}
%
We now describe the quasi-stationary heat transfer under the condition $\alpha\approx 1$ opposite to the adiabatic regime.
For mathematical simplicity, we model the energy injection region with a spherical volume of radius $a$. According to $\alpha \sim 1$, power generation is slow enough, so that the energy outflow becomes
\begin{equation}\label{eq:steadystate} P_{\rm out}=-4\pi a^2\chi|\nabla T|\end{equation}
where $T$ is obtained from the stationary conditions. Denoting $T_a$ the steady state temperature in that region, Eq. (\ref{eq:steadystate}) yields a temperature increase at distance $r>a$,
\begin{equation}\label{eq:deltaT}\delta T(r)=\delta T_aa/r \quad {\rm with}\quad \delta T_a=T_a-T_0\end{equation}
where $T_0$ is the temperature far away from the affected region (possibly at the system interface). As a result, Eq. (\ref{eq:HT}) takes the form,
\begin{equation}\label{eq:TaP}Nk\partial T_a/\partial t=P-4\pi a\chi (T_a-T).\end{equation}

For the voltage source driven regime, using Eq. (\ref{eq:PR}) for $P$ yields a the equation well known in the chemical kinetics, \cite{FK,LL}
\begin{equation}\label{eq:PTOFF}Nk\partial T/\partial t=P_0\exp(-W_{\rm OFF}/kT_a)-4\pi a\chi (T_a-T_0), \end{equation}
which is unstable for temperatures $T_0$ exceeding the critical temperature $T_c$ corresponding to its zero right-hand-side. Indeed, above $T_c$ the heat generation term wins over that of dissipation, the temperature further increases and the process of `thermal explosion' takes place. Using the exponent linearization, \cite{FK} $1/T_a=1/T_0 -(T_a-T_0)/T_0^2$ one can see that $T_c$ can be found from the condition,
\begin{equation}\label{eq:fm} {\cal F}\equiv\frac{P_0T_c^2\exp(-W_{\rm OFF}/kT_c)}{4\pi a\chi W_{\rm OFF}}=[e]^{-1}\end{equation}
where [e] is the base of natural logarithms. For $T>T_c$, the problem reduces to that of adiabatic regime.

From Eq. (\ref{eq:fm}) the thermal explosion temperature $T_c$ is estimated as
\begin{equation}\label{eq:Tc}T_c=\frac{W_{\rm OFF}}{k\ln (P_0[e]/4\pi a\chi W_{\rm OFF})}\end{equation}
in the approximation where the logarithm in denominator is large. Assuming the ballpark values $\chi \sim 1$ W/m$\cdot$K, $a\sim 10$ nm, and $W_{\rm OFF}\sim 1$ eV yields $ln (P_0[e]/4\pi a\chi W_{\rm OFF})\sim 50$ and thus $T_c$ below room temperature. The practical process temperatures are well above the room temperature and thus belong in the adiabatic regime. [If necessary, $T_a$ in subcritical region can be found from Eq. (\ref{eq:PTOFF}).]

The hypothetical case of quasi-stationary RESET process can be analyzed along the same lines with the obvious replacement $\exp(-W_{\rm OFF}/kT_a)\rightarrow \exp(W_{\rm ON}/kT_a)$ in Eq. (\ref{eq:PTOFF}). We will skip such analysis in view of the Wiedemann-France argument in favor of the adiabatic regime in Sec. \ref{sec:ht} above.

\section{Discussion}\label{sec:disc}

The physical content of the above analysis is that heat transfer details are essential for the SET and RESET kinetics. On the formal level that effect is most clearly exhibits itself through the induction times in Eqs. (\ref{eq:tind}), (\ref{eq:tindR}), (\ref{eq:tind1}) and their related process times $\tau _S$, $\tau _R$, and $\tau _S^1$. In particular, the RESET and SET rates exponentially increase for times exceeding the corresponding induction times. The underlying physics is that such times allow significant enough local temperature increase allowing exponentially more efficient structural transformations.

The existence and role of such induction times appear to be practically important and call upon addressing as potential additions to the published experimental work on SET and RESET transition rates. \cite{lee2020,bernard2010,sharma2015,yoo2017} We note another practically important result of this work, which follows from considering the products $\tau _R^2s/2$, $(\tau _S')^2s'$ and $\tau _Sp$ (the latter for the case of $\tau _S\gg\tau _{ind}^S)$, which all present the total energies deposited during the corresponding transitions. All these products turn out to be independent on the corresponding deposition rates $s$, $s'$, and $p$ thus confirming the earlier empirical observations \cite{maestro2015} (so far unexplained) that only the total deposited energies, but not the deposition rates matter. Note however that the latter conclusion would not apply to the case $\tau _S\ll\tau _{ind}^S$, which remains unverified.

Another application of the above results concerns the field dependencies of transition rates \cite{lee2020,bernard2010,sharma2015,yoo2017} sometimes attributed to the field induced nucleation mechanism. \cite{karpov2008} However Eqs. (\ref{eq:tind}), (\ref{eq:tindR}), (\ref{eq:tind1}) and related $\tau _S$, $\tau _R$, and $\tau _S^1$ predict another source of the transition rates field dependencies as originating from the corresponding induction and process times through the Poole-Frenkel effect stating that the activation energies ($W_{\rm OFF}$ and $W_{\rm ON}$) decrease with the field strength. \cite{schroeder2015} Indeed, it follows from the above results that induction and process times will decrease following the Poole-Frenkel exponents thus accelerating the transition rates. The underlying physics is related to the heat transfer as well: suppressing the conductivity barrier increases Joule heat. It can be verified experimentally by correlating the Poole-Frenkel vs. transition rates field dependencies. Note that the latter explanation does not rule out the field induced nucleation mechanism just adding some alternatives in parallel.

It is natural to hypothesize that here presented theory can be extended beyond the domain of memory SET/RESET transitions towards more general problems of dielectric breakdown and operations of electric fuses. The former could generalize the above described SET process to various other conditions making two basic predictions: the figure of merit of Eq. (\ref{eq:fm}) for the breakdown conditions, and the induction/process development times. The latter would extend the above RESET description in Eqs. (\ref{eq:rate0}) - (\ref{eq:tR}) predicting the fuse breaking induction time and kinetics vs. ambient temperature and fuse dimensions. Such extensions remain to be developed and validated by comparing against a significant body of the available experimental data.

We shall end this section with a remark establishing connections with a practically important problem of energy consumption. It was empirically found that the RESET switching current is proportional to the device electrode area $S$ pointing towards ultimate decrease of the latter for energy minimization. \cite{raoux2014} We note however that such a relation between the current and electrode area is inconsistent with the commonly shared picture of the current flowing through a narrow conducting filament whose diameter is unrelated to $S$. The inconsistency can be reconciled by interpreting the observed proportionality as related to the displacement current $I=I_D=(S/4\pi) \partial D/\partial t$ (where $D$ is the electric induction and we use Gaussian system of units). The corresponding power $P_D=I_DEl=(\partial /\partial t)(Sl ED/8\pi)$ (with $E$ being the field strength and $l$ the interelectrode distance) represents the full time derivative of the electrostatic energy that is not dissipated and returns to the power source after the pulse is completed; hence, no electrode area concerns. Note that by the same token, our consideration above neglected the displacement current component in describing the heat transfer problem.

\section{Conclusions}\label{sec:concl}
We have developed a simple analytical description of SET and RESET transitions in solid state memory devices. Our approach is based on the adiabatic approximation adequate for practical device parameters and conditions.

Our theory partially utilized  a remarkable formal similarity between the SET processes and the established patterns of thermal explosion in chemical kinetics. \cite{FK} That similarity can be further explored in the future.

Finally, we note a number of predictions that can trigger a related experimental research, such as the induction times in SET and RESET transitions, and correlations between Poole-Frenkel and switching rate electric field  exponents.

The data that supports the findings of this study are available within the article.

\end{document}